\documentclass[%
aip,
amsmath,amssymb,
reprint,%
floatfix,
]{revtex4-1}

\usepackage{graphicx}
\usepackage{dcolumn}
\usepackage{bm}

\usepackage{amssymb}
\usepackage{color}

\newcommand{\umux}{$\mu$mux} 

\begin{document}
\title{A Microwave SQUID Multiplexer Optimized for Bolometric Applications}
\author{B.~Dober}
\email{bradley.dober@nist.gov \\ Contribution of the National Institute of Standards and Technology; not subject to copyright in the United States}
\affiliation{Department of Physics, University of Colorado Boulder, Boulder, CO 80309, USA}
\affiliation{National Institute of Standards and Technology, 325 Broadway, Boulder, CO 80305, USA}
\author{Z.~Ahmed}
\affiliation{Kavli Institute for Particle Astrophysics and Cosmology, SLAC National Accelerator Laboratory, Menlo Park, CA 94025, USA}
\affiliation{SLAC National Accelerator Laboratory, Menlo Park, CA 94025, USA}
\author{K.~Arnold}
\affiliation{Department of Physics, University of California San Diego, La Jolla, CA 92093, USA}
\author{D.T.~Becker}
\affiliation{Department of Physics, University of Colorado Boulder, Boulder, CO 80309, USA}
\affiliation{National Institute of Standards and Technology, 325 Broadway, Boulder, CO 80305, USA}
\author{D.A.~Bennett}
\affiliation{National Institute of Standards and Technology, 325 Broadway, Boulder, CO 80305, USA}
\author{J.A.~Connors}
\affiliation{Department of Physics, University of Colorado Boulder, Boulder, CO 80309, USA}
\affiliation{National Institute of Standards and Technology, 325 Broadway, Boulder, CO 80305, USA}
\author{A.~Cukierman}
\affiliation{Kavli Institute for Particle Astrophysics and Cosmology, SLAC National Accelerator Laboratory, Menlo Park, CA 94025, USA}
\affiliation{Department of Physics, Stanford University, Stanford, CA 94305, USA}
\author{J.M.~D'Ewart}
\affiliation{SLAC National Accelerator Laboratory, Menlo Park, CA 94025, USA}
\author{S.M.~Duff}
\affiliation{National Institute of Standards and Technology, 325 Broadway, Boulder, CO 80305, USA}
\author{J.E.~Dusatko}
\affiliation{SLAC National Accelerator Laboratory, Menlo Park, CA 94025, USA}
\author{J.C.~Frisch}
\affiliation{SLAC National Accelerator Laboratory, Menlo Park, CA 94025, USA}
\author{J.D.~Gard}
\affiliation{Department of Physics, University of Colorado Boulder, Boulder, CO 80309, USA}
\author{S.W.~Henderson}
\affiliation{Kavli Institute for Particle Astrophysics and Cosmology, SLAC National Accelerator Laboratory, Menlo Park, CA 94025, USA}
\affiliation{SLAC National Accelerator Laboratory, Menlo Park, CA 94025, USA}
\author{R.~Herbst}
\affiliation{SLAC National Accelerator Laboratory, Menlo Park, CA 94025, USA}
\author{G.C.~Hilton}
\affiliation{National Institute of Standards and Technology, 325 Broadway, Boulder, CO 80305, USA}
\author{J.~Hubmayr}
\affiliation{National Institute of Standards and Technology, 325 Broadway, Boulder, CO 80305, USA}
\author{Y.~Li}
\affiliation{Department of Physics, Princeton University, Princeton, NJ 08544, USA}
\author{J.A.B.~Mates}
\affiliation{National Institute of Standards and Technology, 325 Broadway, Boulder, CO 80305, USA}
\author{H.~McCarrick}
\affiliation{Department of Physics, Princeton University, Princeton, NJ 08544, USA}
\author{C.D~Reintsema}
\affiliation{National Institute of Standards and Technology, 325 Broadway, Boulder, CO 80305, USA}
\author{M.~Silva-Feaver}
\affiliation{Department of Physics, University of California San Diego, La Jolla, CA 92093, USA}
\author{L.~Ruckman}
\affiliation{SLAC National Accelerator Laboratory, Menlo Park, CA 94025, USA}
\author{J.N.~Ullom}
\affiliation{National Institute of Standards and Technology, 325 Broadway, Boulder, CO 80305, USA}
\author{L.R.~Vale}
\affiliation{National Institute of Standards and Technology, 325 Broadway, Boulder, CO 80305, USA}
\author{D.D.~Van Winkle}
\affiliation{SLAC National Accelerator Laboratory, Menlo Park, CA 94025, USA}
\author{J.~Vasquez}
\affiliation{SLAC National Accelerator Laboratory, Menlo Park, CA 94025, USA}
\author{Y.~Wang}
\affiliation{Department of Physics, Princeton University, Princeton, NJ 08544, USA}
\author{E.~Young}
\affiliation{Department of Physics, Stanford University, Stanford, CA 94305, USA}
\affiliation{Kavli Institute for Particle Astrophysics and Cosmology, SLAC National Accelerator Laboratory, Menlo Park, CA 94025, USA}
\author{C.~Yu}
\affiliation{Department of Physics, Stanford University, Stanford, CA 94305, USA}
\author{K.~Zheng}
\affiliation{Department of Physics, Princeton University, Princeton, NJ 08544, USA}

\begin{abstract}

A microwave SQUID multiplexer (\umux) has been optimized for reading out large arrays of superconducting transition-edge sensor (TES) bolometers. 
We present the scalable cryogenic multiplexer chip design that may be used to construct an 1820-channel multiplexer for the 4-8~GHz rf band.
The key metrics of yield, sensitivity, and crosstalk are determined through measurements of 455 readout channels, which span 4-5~GHz.  
The median white-noise level is 45~pA/$\sqrt{\textrm{Hz}}$, evaluated at 2~Hz, with a 1/f knee $\leq$~20~mHz after common-mode subtraction.  
The white-noise level decreases the sensitivity of a TES bolometer optimized for detection of the cosmic microwave background at 150~GHz by only 3\%.  
The measured crosstalk between any channel pair is $\leq$~0.3\%.  

\end{abstract}

\maketitle




For many scientific applications involving photon-sensing low-temperature detectors, measurement sensitivity is limited by fluctuations intrinsic to the signal of interest.    
As such, experiments implement arrays of photon-noise-limited sensors to improve signal-to-noise ratio.  
Array size is limited by the ability to combine signals into a manageable number of output wires to transmit signals from the cryogenic stage to the room temperature readout electronics.  
For power-sensing instruments based on transition-edge-sensors (TESs), time-division multiplexing (TDM) \cite{TDMmux} and megahertz frequency-division multiplexing (FDM) \cite{FDMmux,jackson2011spica} are well-established techniques, which to date have been used in fielded experiments to combine a maximum of 68 sensors into one wiring/amplification chain \cite{spt3g,henderson2016}.
The many instruments that require thousands to hundreds of thousands of bolometers \cite{simonsdecadal,alicpt,cmbs4,vavagiakis2018ccatprime,biceparray,leisawitz2018ost} stretch the capability of these established techniques, which are fundamentally limited by their $\sim$10~MHz output channel bandwidth.  
To increase the number of sensors per wiring/amplification chain (here referred to as the multiplexing factor) and enable more sensitive bolometric arrays, multiplexing techniques that make use of the microwave readout band are under development.  
These techniques include microwave kinetic inductance detectors (MKIDs) \cite{day2003}, kinetic inductance parametric upconverters (KPUPs) \cite{kher2016kpup}, and the subject of this letter, the microwave superconducting quantum interference device (SQUID) multiplexer (\umux)\cite{irwinlehnert2004,jabmates,matesumux2008,kempf2014multiplexed,hirayama2013microwave}. 
Application to calorimetric instruments has been previously described \cite{mates2017sledgehammer,bennett2019microwave}.  
Here we focus on bolometric applications, with a particular emphasis on measurements of the cosmic microwave background (CMB).  
Preliminary work has demonstrated the feasibility of the \umux\ for CMB measurements \cite{dober2017apl}. 
The MUSTANG2\cite{stanchfield2016} 90~GHz receiver coupled to the Greenbank Telescope (GBT) operates using a 4$\times$64 channel \umux\ readout. 
In addition, a Keck Array receiver has been retrofitted with an 8$\times$64 channel \umux\ readout and has spent a season observing the CMB at 150~GHz\cite{cukierman2020microwave}.
In this letter, we present a \umux100k multiplexer chip that has been optimized for the readout of TES bolometers.
Multiple frequency-scaled versions of this chip can be combined to form a nearly 2000 sensor multiplexer within one octave of rf bandwidth. 

The principle of operation of the \umux\ has been described in previous publications \cite{jabmates,irwinlehnert2004,mates2012flux,mates2017sledgehammer}.  
Briefly, an rf-SQUID transduces a dc-biased TES signal into the frequency shift of an approximately gigahertz quarter-wave resonator. 
Each TES is coupled to its own SQUID-coupled resonator that has a unique resonant frequency. 
All resonances are coupled to a common co-planar waveguide (CPW) microwave readout line.  
An additional source of SQUID flux is ramped to linearize the SQUID response under the paradigm of flux-ramp modulation\cite{mates2012flux}.

The \umux100k multiplexer represents a significant departure from previous work \cite{jabmates,dober2017apl,mates2017sledgehammer} and has been influenced by its application to the Simons Observatory\cite{simonsdecadal}, a set of CMB imagers sensitive to a broad range of angular scales.
In general, the multiplexer architecture achieves an 1820 multiplexing factor within a 4-8~GHz readout band.  
All components are designed to fit within a two-dimensional plane behind a 150~mm detector wafer, which aids in tiling multiple wafers into a single focal plane \cite{sopackaging}.  
The readout contributes $<$~5\% to the overall noise of the detectors, which translates to an input-referred current noise of 45~pA/$\sqrt{\mathrm{Hz}}$.  
The maximum cross-talk between readout channels is $<$~0.3\%.  
Lastly, the absolute frequency placement of the resonators is designed to match the usable rf bandwidth of the room temperature electronics\cite{henderson2018smurf}.

\begin{figure}[t]
    \centering
    \includegraphics[width=1.0\linewidth]{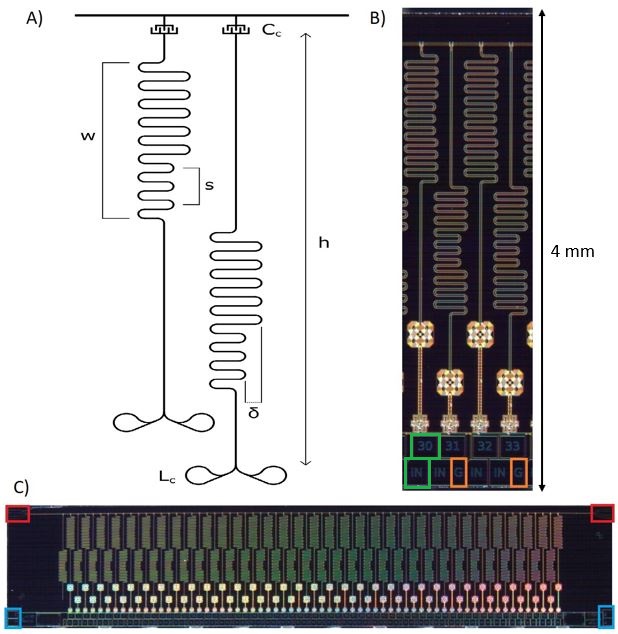}
    \caption{ \umux~chip overview and resonator cells. A: a schematic of the key features of the \umux~resonator definition. A $5\times10\times5~\mathrm{\mu m}$ CPW feedline runs along the top and is coupled to the resonators using an interdigitated capacitive (IDC) coupler, which defines the coupling capacitance ($C_{c}$). Below is the alternating resonator meander, whose total length (h) is set by the number of meanders (w), number of sliders (s), and length of sliders ($\delta$). The sliders are used by the lithographic stepper to set the unique resonant frequency of each readout channel. $L_{c}$ represents the effective self-inductance of the coupled rf SQUID. B: an optical micrograph of several \umux~channels.  The TES inputs (green squares) are connected using the numbered bond pads shown at the bottom. The additional narrower bond pad is used to tie the ground plane of the chip to the packaging. C: An optical micrograph of an entire 4$\times$20~mm~ \umux~chip. The CPW feedline runs along the top (red squares). Along the bottom, there are input bond pads for the flux ramp on both sides of the chip (blue squares), as well as bond pads for the TES inputs and for grounding to the packaging (orange squares in B). 
    }
    \label{fig:design}
\end{figure}


Fig.~\ref{fig:design} shows the \umux100k multiplexer chip, which satisfies the criteria stated in the previous paragraph.  
Each 4$\times$20~mm$^2$ chip has 65 readout channels plus one resonator without a SQUID.  
Multiple frequency-scaled versions of the \umux100k chip may be connected in series (``daisy-chained") via aluminum wirebonds to create a larger multiplexer.    
The resonator without a SQUID is intended to track the two-level system noise (TLS) of the resonators. The user may opt to leave one of the 65 readout channels disconnected to track both the readout noise of the system, as well as magnetic pickup in the SQUIDs.

\begin{table}
\caption{\umux100k specifications}
\label{table:spec}
\begin{tabular}{ |l|c|c|  }
 \hline
 Parameter & Symbol & Value \\
 \hline
 Die size & & 4x20~mm$^{2}$ \rule{0pt}{2.6ex} \\
 Channels per die & $N$ & 65 \\
 Resonator bandwidth & $BW$ & 100~kHz \\
 Resonant frequency & $f_o$ & 4-8~GHz \\
 Minimum frequency spacing & $\Delta{f}$ & 1.8~MHz \\
 Input mutual inductance & $M_{in}$ & 227~pH \\
 Flux ramp mutual inductance & $M_{fr}$ & 13.3~pH \\
 Resonator mutual inductance & $M_{res}$ & $\sim$1.3~pH \\
 Frequency shift & $df_{pp}$ & 100~kHz \\
 \hline
\end{tabular}
\end{table}

Table~\ref{table:spec} summarizes the chip specifications.    
A key distinction relative to previous multiplexers is the reduced resonator bandwidth ($BW$~=~100~kHz), which lends the multiplexer version its name (\umux100k).  
The $BW$ is controlled by adjusting the capacitive coupling ($Q_{c}$) to the CPW feedline and has been chosen to maximize the number of channels within one octave of readout bandwidth while considering both the signal bandwidth and the flux ramp rate needed to overcome sources of 1/f noise. $Q_{c}$ is set by varying the length of the three fingers which define the interdigitated capacitor (IDC).
A four-lobe gradiometric SQUID\cite{gradiometricsquid} is utilized to reduce sensitivity to external magnetic fields. The TES input mutual inductance ($M_{in}$) is set by a loop winding through all four lobes, while the flux ramp mutual inductance ($M_{fr}$) is set by a smaller loop winding through the lower two SQUID lobes.
The resonant frequency ($f_o$) is periodic with applied flux, and the peak-to-peak frequency swing ($df_{pp}$) is set to match the resonator $BW$ by tuning the SQUID-to-resonator coupling ($M_{res}$). This is tuned by varying the size of a coil which winds through the upper two SQUID lobes.

The resonant frequencies and spatial resonator layout are designed to achieve high spatial density while suppressing several sources of cross-talk.    
The design principle is to distribute resonators such that spatial neighbors are largely separated in frequency space, and frequency neighbors are largely separated in spatial distance \cite{mates2019crosstalk}.  
We include several frequency gaps that guard against frequency collisions, which may arise due to intra-wafer variation. 
Additionally, there are 38~MHz wide frequency gaps every 500~MHz to accommodate the input quadruplexers of the SLAC Microresonator Radio Frequency (SMuRF) room-temperature electronics \cite{henderson2018smurf}.  
These choices lead to a non-uniform resonator frequency schedule, which repeats every 500~MHz and spans 3.5 chips. 
Each of the 66 channels comprising a single multiplexer band are split into two halves and placed into upper and lower rows of resonators to maximize spatial density of the channels (see Fig.~\ref{fig:design}).   
In each half band, the 33 channels are grouped into two additional sub-bands of 17 and 16 channels that are interleaved in spatial distance on the chip. 
Within a sub-band, resonators are spaced by $\Delta{f}$~=1.8~MHz, the minimum designed frequency spacing. 
Between sub-bands, half-bands, and bands, there are additional fixed gaps of 3.06~MHz, 4.5~MHz, and 6.3~MHz, respectively. 
This grouping results in either a 32.0~MHz or 65.7~MHz space between nearest spatial neighbors and $>$ 31.7~MHz for next-nearest neighbors.  
Frequency adjacent channels are spaced 1~mm apart.
With this configuration, we simulate in Microwave Office a maximum cross-talk of 0.2\%, dominated by nearest-frequency-neighbor resonator-to-resonator cross-talk. Other sources of cross-talk contribute less than $<$0.1\%.

As with all FDM systems, resonator frequency collisions or omissions complicate mapping resonant frequencies to optical pixels, which is required in most instruments.  The chips measured in this letter demonstrate no collisions or resonant frequency swapping, and only four missing resonators. The limited number of missing resonators, coupled with the resonance grouping technique outlined here limits these concerns to a manageable level for the demonstrated frequency density.  Full screening of multiplexer chips before integration in larger instruments can also aid frequency to pixel mapping.

Device fabrication largely follows the description in the work of Mates\cite{jabmates}.
In brief, these devices are fabricated on 3 inch diameter high-resistivity silicon wafers which are covered with a minimal layer (20~nm) of $\mathrm{SiO}_{\mathrm{2}}$ in an effort to reduce its TLS noise contribution. First, the Josephson junction process begins with depositing a trilayer of niobium (200~nm), aluminum ($\sim$ 7~nm) which is partially oxidized to form the insulating barrier, and niobium (120~nm). The top two layers are etched away to form the 2.5 $\times$ 2.5~$\mu m^{2}$ junction pillars, with the bottom constituting the first wiring layer where the majority of the circuitry is defined. Next, a $\mathrm{SiO}_{\mathrm{2}}$ insulating layer (350~nm) is deposited. Etching holes through this layer allows for the creation of vias. An additional niobium layer (300~nm) is deposited to connect the junctions to the first wiring layer and to create the CPW feedline ground-straps. In the penultimate step all $\mathrm{SiO}_{\mathrm{2}}$ is etched away wherever possible to reduce TLS noise. The CPW resonators are etched in the final step of fabrication so that no other process can contaminate the resonant cavity.
To efficiently microfabricate $\sim$2000 unique resonant frequency cells, we employ a lithographic stepper-based fabrication technique, similar to the tile-and-trim approach for fabricating MKIDs\cite{mckenney2019tile}. 
With reference to the schematic in Fig.~\ref{fig:design}, all CPW resonators within a half-band are flashed by a single image that consists of $w$ CPW meander turns and $s$ unexposed turns.
Resonant frequencies are subsequently defined by shooting a second image, which completes the $s$ turns with a reduced turn length $\delta$, that realizes a unique CPW length.      
There is a factor of $\sim$~6 reduction in nearest-neighbor frequency scatter when employing this technique relative to shooting all resonators on a chip with a single, large mask.  
Additionally, re-configuring the frequency schedule to optimize from one fabrication round to the next, or even to meet the needs of an entirely different experiment, requires only a new stepper job file.  
New mask generation is not required.



\begin{figure}
    \centering
    \includegraphics[width=1.0\linewidth]{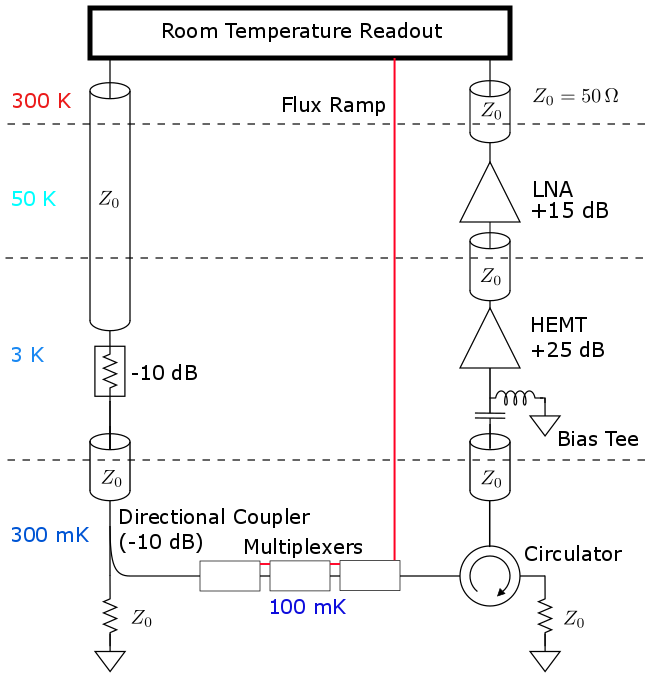}
    \caption{ \umux~rf wiring schematic. The input microwave tones are attenuated 10~dB at both the 3~K and 300~mK stages via a fixed attenuator and a directional coupler, respectively, before entering the \umux\ at the 100~mK stage. On the output, after passing through a circulator and a bias tee, the modulated tone is amplified by a +25~dB HEMT at 4~K and a +15~dB low noise amplifier (LNA) at 50~K. The two-stage low gain amplifier chain has higher 3~dB compression point than a single 4~K LNA of the same gain, allowing for a larger number of readout channels before saturating the amplifiers.
    }
    \label{fig:schem}
\end{figure}

To test this architecture, we assembled a seven-chip multiplexer spanning 4-5~GHz into a copper device box\cite{mates2017sledgehammer} and installed the package into a $\textrm{He}^{3}$-backed adiabatic demagnetization refrigerator (ADR) cooled to 100~mK.  
The experimental setup detailing the rf wiring is shown in Fig.~\ref{fig:schem}.
We used a commercial vector network analyzer (VNA) for microwave transmission measurements.
For noise and cross-talk measurements, we operated the SMuRF room-temperature electronics with a 20~kHz SQUID modulation rate and utilized resonator tone-tracking\cite{henderson2018smurf}.  
This resulted in a 4~kHz effective sampling rate, which we further down-sampled to 200~Hz.   
Resonances were interrogated with microwave probe tone powers near -73~dBm (referred to the input of the multiplexer chip feedline), which is near optimal for noise performance.  

\begin{figure}
    \centering
    \includegraphics[width=1.0\linewidth]{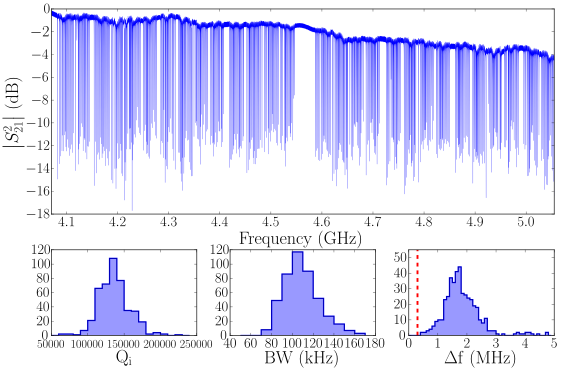}
    \caption{
    Transmission and resonator parameters for a seven-chip multiplexer spanning 4-5~GHz. Top: $ \left| S_{21}^2 \right| $ as a function of frequency. Bottom: histograms of $Q_{i}$, BW, and frequency spacing ($\Delta f$). Counts to the left of the red dashed line indicate notional frequency collisions ($\Delta{f} <$ 3BW), of which there are zero.
 }
    \label{fig:specs}
\end{figure}

Fig. \ref{fig:specs} presents a frequency survey and the channel statistics based on these data.
We fit the complex transmission of each resonance to a model\cite{gaothesis} and determine the resonator parameters $f_o$, internal quality factor ($Q_{i}$), and $BW$.  
We identified 458 resonances out of a possible 462.  
The mean resonator spacing is 1.9~MHz, which is close to the design goal.  
All resonator pairs are separated by $>$~3$\times BW$, which we deem collision-free.
The mean $Q_{i}$ = 128,348 is generally consistent with our experience in multiple rounds of \umux100k fabrication.  
The mean $BW$=115~kHz is 15\% higher than the designed value and is a result of the coupling quality factors ($Q_{c}$) being chosen with the assumption that $Q_{i} = 200,000$.
In summary, we expect $>$~99\% initial multiplexer channel yield from these measurements, which is typical of these devices.


\begin{figure}
    \centering
    \includegraphics[width=1.0\linewidth]{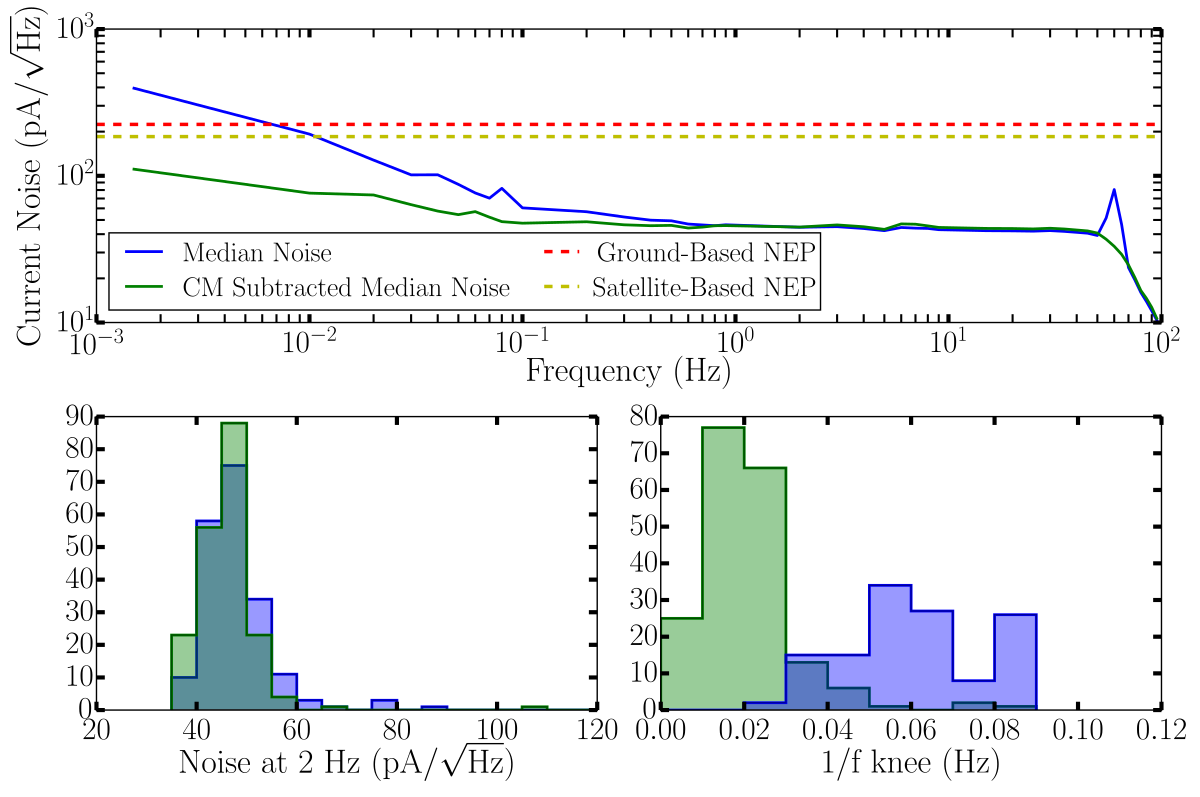}
    \caption{\umux100k noise.  
    Top: Median current noise without (blue, solid line) and with (green, solid line) common-mode subtraction of 195 channels.   Red (yellow) dashed lines are the expected photon-noise levels of a 150~GHz ground-based (space-based) CMB detector, which are a factor of 5 (4) higher than the measured multiplexer white-noise level.    
    Bottom: Histograms of the noise at 2 Hz (left) and 1/f knee (right) without (blue) and with (green) common-mode subtraction. 
    }
    \label{fig:noise}
\end{figure}

To determine the noise performance of the multiplexer, the TES inputs were left open and 800~s of data were simultaneously streamed from the channels within the first 500~MHz of rf bandwidth.  
Of the possible 227 channels, the electronics successfully tone-tracked 205, for a yield of 90.3$\%$.  
The remaining 22 channels were disabled by the SMuRF due to improper automatic resonator calibration. Several of these channels may be recoverable with better resonator tuning parameters.
A linear drift subtraction was the only time-domain data processing step.
For each resonator, we compute the amplitude power spectral density using multiple Welch periodograms at several frequency resolutions.  
The SMuRF tracking algorithm natively returns the flux-ramp demodulated phase ($\phi$) in radians.   
We convert the demodulated phase noise ($S_{\phi}$) to input current noise ($S_{I}$) (or equivalently noise equivalent current (NEI)) by use of the relation
\begin{equation}
    \label{eqn:phi2I}
    S_{I} = \frac{2\pi\Phi_o}{M_{in}} S_{\phi},
\end{equation}
where $\Phi_o$ is the magnetic flux quantum and $M_{in}$ is the mutual inductance between the rf SQUID and the TES.
Fig.~\ref{fig:noise} presents the results.  
The median white-noise level is 45~pA/$\sqrt{\mathrm{Hz}}$.
The 1/f knee, defined as the frequency at which the noise is twice the white-noise level (evaluated between 50-100~Hz), is 64~$\pm$~31~mHz.  
When subtracting one readout channel's timestream from all others (a naive form of common-mode subtraction) and computing the power spectral densities, the peak in the 1/f knee histogram reduces to $\sim$~20~mHz.  
The true value may be lower still because measurement is limited by the 800~s measurement time.  
We note the naive common-mode subtraction is valid in the limit that all channels have the same gain.  
Eqn.~\ref{eqn:phi2I} shows that $M_{in}$ is the single parameter that governs the gain.  
Variation of this parameter is geometry-dependent and set by micro-lithography. We simulate via FastHenry the maximum over-etch possible during lithography (100~nm), which produces $<$0.8\% deviation from the designed value.
The absence of the 60~Hz line in the common-mode subtracted power spectrum suggests that this assumption is true.

These measured noise characteristics are highly favorable for measurements of the cosmic microwave background.  
The expected photon noise of a 150~GHz channel in a ground-based\cite{cmbs4} CMB experiment with a 3.10~pW photon load is 30.5~aW/$\sqrt{\mathrm{Hz}}$. Similarly, the expected photon noise for a satellite-based\cite{sugai2020litebird} experiment with a 0.46~pW load is 9.77~aW/$\sqrt{\mathrm{Hz}}$. 
To put the noise equivalent power (NEP) of these experiments in context of the current noise of the multiplexer, we use the following equation
\begin{equation}
    \label{eqn:NEP2NEI}
    S_{I} = \frac{NEP}{\sqrt{P_eR_o}},
\end{equation}
which assumes an ideal bolometer in the high-loop gain limit with the electrical power ($P_e$) equal to 1.5~times the optical power and the bolometer operating resistance $R_o$~=~4~m$\Omega$. Significant deviations in sensor resistance from 4~m$\Omega$ may require changing $M_{in}$ in order to maintain the stated performance.
The measured multiplexer white-noise level is one fifth and one fourth of the expected current noise for the ground and satellite-based experiments. Therefore, the multiplexer decreases the sensitivity to photon noise by only 2\% and 3\% respectively for the ground and satellite cases.
Furthermore, the low 1/f knee provides access to large angular scale measurements, which are required for CMB B-mode polarization searches.

\begin{figure}
    \centering
    \includegraphics[width=1.0\linewidth]{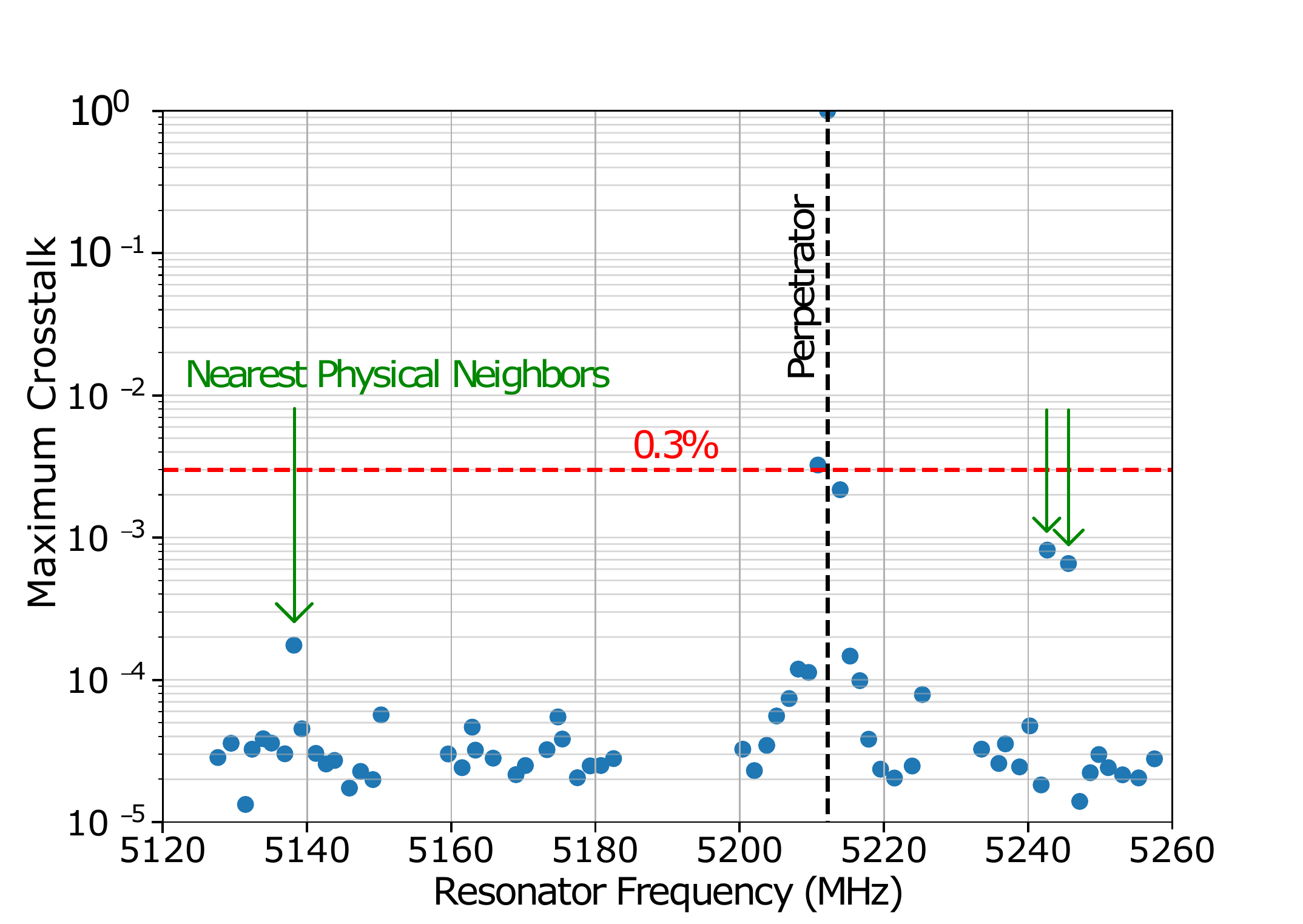}
    \caption{ 
    Maximum \umux100k channel cross-talk plotted versus resonant frequency.  
    Vertical black, dashed line indicates the position of the perpetrator channel.  
    Cross-talk is universally $\leq$~0.3\%, with the highest offenders from nearest frequency neighbors.   
    Green arrows indicate that nearest spatial neighbors' crosstalk at $<$~0.1\%.}
    \label{fig:crosstalk}
\end{figure}

As in any multiplexer, combining multiple signals into one wiring/amplification chain may lead to unwanted sources of cross-talk between signal channels.  
Sources of cross-talk particular to the microwave SQUID multiplexer are discussed by Mates et al.\cite{mates2019crosstalk}.
To quantify the cross-talk of the \umux100k multiplexer, a single chip was packaged in a device box and installed in a separate ADR cryostat with rf wiring similar to that shown in Fig.~\ref{fig:schem} and read out with tone-tracking via SMuRF.  
The sum of a dc and sinusoidal current was injected into a single channel's input (referred to as the ``perpetrator" channel), and the response of the 64 ``victim" channels was observed. 
The amplitude of the sinusoidal signal was chosen to produce $\sim \Phi_0/10$, so as to measure the differential crosstalk at a single dc current level.  
The fractional crosstalk response for a given channel was calculated using a lock-in demodulation technique relative to the perpetrator response.  
The measurement was repeated as the dc current level was stepped in $\Phi_0/10$ intervals across several $\Phi_0$, as crosstalk arising from resonator hybridization is expected to vary sinusoidally with the SQUID dc flux offset\cite{mates2019crosstalk}. 
The largest fractional cross-talk across all dc current levels for each channel is reported in Fig.~\ref{fig:crosstalk}. 
The highest measured cross-talk was 0.3\% which corresponds to channels closest in frequency to the perpetrator.  
Spatial neighbors of the perpetrator channel display the next-highest level of cross-talk ($<0.1\%$), and all other victim channels show cross-talk at or below one part in $10^{4}$.
While these results come from measurements on a single chip, we expect the results to hold for higher channel-count multiplexers with the exception of cross-talk that results from intermodulation products, which scale with the number of microwave tones.  Tone-tracking and careful selection of linear amplifiers ameliorate this source of cross-talk.
These results are less than or equal to the cross-talk in TDM systems\cite{TDMmux} that have been deployed in tens of TES-based instruments.   

Large-sensor-count bolometric experiments pose a significant readout challenge.  
To meet these demands, we developed the \umux100k multiplexer.   
We have presented the \umux100k chip design as well as the topology to construct a 1820-channel multiplexer within the 4-8~GHz readout band.  
Key metrics of yield, noise, and cross-talk have been quantified on resonators that span more than the fundamental repeating unit of the multiplexer, and these metrics meet or exceed the requirements of large-scale bolometric instruments currently under development.  
The multiplexer is scalable in sensor count and rf bandwidth, and flexible in defining a resonant frequency schedule.  
As such, the design may be tailored to other applications in a straightforward manner.  

\begin{acknowledgments}
The authors acknowledge the support of the Simons Foundation  (Award 457687, B.K.), the NIST Innovations in Measurement Science program, and the NASA APRA program. The effort at SLAC was supported by the Department of Energy, Contract DE-AC02-76SF00515. 

The data that support the findings of this study are available from the corresponding author
upon reasonable request.
\end{acknowledgments}

\bibliography{citations}

\end{document}